\documentclass[aps,prd,twocolumn,superscriptaddress]{revtex4}
\usepackage{graphicx}
\usepackage{slashed}
\usepackage{ragged2e}
\usepackage{caption}
\DeclareCaptionJustification{justified}{\justifying}
\usepackage{subcaption}
\usepackage{float}
\usepackage{amsmath}
\usepackage{amsthm}
\usepackage{amsfonts}
\usepackage{amssymb}
\usepackage{dsfont}
\usepackage{multirow}
\usepackage{appendix}
\usepackage{tikz}
\usetikzlibrary{shapes, arrows,positioning, calc}
\usepackage{lipsum}
\usepackage{hyperref}
\hypersetup{
	colorlinks=true,
	linkcolor=blue,
	filecolor=magenta,      
	urlcolor=cyan,
	citecolor=blue,
}
\usepackage{titlesec}
\titlespacing*{\section}{0pt}{12pt}{6pt} % {sol}{üst}{alt}
\begin{document}
	
\title{Temperature Effects on a Vector Hidden-Charm Molecule}
\author{E.~G\"ung\"or}
\affiliation{Department of Physics, Kocaeli University, 41001 Kocaeli, T\"urkiye}
\author{H.~Sundu}
\affiliation{Department of Physics Engineering, Istanbul Medeniyet University, 34700 Istanbul, T\"urkiye}
\author{J.Y.~S\"{u}ng\"{u}}
\affiliation{Department of Physics, Kocaeli University, 41001 Kocaeli, T\"urkiye}
\author{E. Veli Veliev}
\affiliation{Department of Physics, Kocaeli University, 41001 Kocaeli, T\"urkiye}
\date{\today}

\begin{abstract}
We investigate the thermal properties of the $Y(4500)$ state within the framework of 
thermal QCD sum rules, assuming a $D_s \bar{D}_{s1}$ molecular configuration with 
$J^{PC}=1^{--}$. The analysis is performed at both zero and finite temperatures, 
employing the operator product expansion up to dimension-5 condensates. The Borel 
window and continuum threshold are carefully selected to ensure OPE convergence and 
pole dominance. As the temperature approaches the deconfinement temperature $T_c$, 
the $Y(4500)$ undergoes significant medium modifications: its mass decreases by $29\%$ 
and its decay constant is suppressed by $94\%$ relative to their vacuum values, while 
the decay width increases by $35\%$, signaling the dissociation of the state in the 
medium. These results indicate that the $Y(4500)$ becomes unstable near $T_c \approx 
155~\mathrm{MeV}$, consistent with its melting into the quark-gluon plasma. The 
obtained thermal spectral parameters may serve as signatures for identifying the 
$Y(4500)$ in heavy-ion experiments at RHIC and LHC, and provide predictions for 
sequential suppression patterns in the exotic hadron sector.
\end{abstract}

\maketitle

%%%%%%%%%%%%%%%%%%%%%%%%%%%%%%%%%%%%%%%%%%%%%%%%%%%
\section{Introduction}\label{sec:intro}
%%%%%%%%%%%%%%%%%%%%%%%%%%%%%%%%%%%%%%%%%%%%%%%%%%%
The last twenty years have witnessed the experimental emergence of various exotic hadrons that defy traditional quark model frameworks~\cite{Chen:2022asf,Chen:2016qju,Zhu:2024swp,Brambilla:2019esw,Ali:2017jda,Zhou:2025yjb,Tan:2025nir,Wang:2025apq}. In particular, in the charmonium sector, these states have attracted significant theoretical attention concerning their internal structure and formation mechanisms~\cite{Chen2016,Esposito2017,Olsen2018,Wang:2013exa}. Among these, the $Y(4500)$ state was observed by the BESIII Collaboration in the 
$e^+e^- \to K^+K^- J/\psi$ process. It was assigned the quantum numbers 
$J^{PC} = 1^{--}$, with a measured mass of 
$M=(4484.7\pm13.3\pm24.1)~\mathrm{MeV}$ 
and a width of 
$\Gamma=(111.1\pm30.1\pm15.2)~\mathrm{MeV}$~\cite{BESIII:2022joj}.

The nature of this state remains under debate, with theoretical interpretations ranging from conventional charmonium states to molecular configurations or tetraquark structures. In this context, Wang and Liu~\cite{Wang:2022jxj} predict that $Y(4500)$ is a higher $5S$-$4D$ mixed charmonium state $\psi(4500)$, supported by the newly observed enhancement in $e^+e^- \to K^+K^-J/\psi$. The observed $Y(4500)$ mass aligns well with the predicted $\psi(4500)$, though the width discrepancy ($2\sigma$ larger) was resolved by considering interference effects and additional resonance contributions. In other work~\cite{Peng:2022nrj}, the $Y(4500)$ state is interpreted as a vector hadronic molecule composed of a $D_s \bar{D}_{s1}$ meson pair within the framework of heavy-quark spin symmetry and SU(3) flavor symmetry. In our previous work~\cite{Gungor:2023ksu}, incorporating QCD condensates up to operator dimension ten, we estimated the mass of $Y(4500)$ to be $(4488.35 \pm 11.54)$ MeV and the decay constant $f_Y = (4.04 \pm 0.36)\times 10^{-3}~\mathrm{GeV^4}$ at zero temperature. 

Furthermore, the behavior of conventional and exotic hadrons in a thermal medium is crucial for understanding the phase transition from hadronic matter to quark-gluon plasma (QGP) in heavy-ion collisions~\cite{Rafelski:2015cxa}. The QGP represents a state of matter where quarks and gluons are no longer confined within hadrons but exist as free particles in a hot, dense medium. This phase is believed to have existed microseconds after the Big Bang and can be recreated in high-energy heavy-ion collisions at facilities such as RHIC~\cite{Gyulassy:2004zy,Busza:2025uid} and LHC~\cite{ALICE:2010suc}. Numerous studies in the literature have extensively explored the properties, production mechanisms, and decay channels of exotic and conventional particles under high-temperature conditions, particularly in the context of quark–gluon plasma formation in heavy-ion collisions~\cite{Zhang:2025fcv,Torres-Rincon:2024sah,Aydin:2025lbl,Sungu:2020zvk, Azizi:2019kzj,Zhao:2020nwy,Dominguez:2007ic,Turkan:2022gho,Turkan:2020lfo,Llanes-Estrada:2019wmz,Sungu:2020azn,Turkan:2019anj}.

Exotic hadrons like $Y(4500)$ are particularly sensitive to medium effects due to their complex internal structures and weak binding energies, making them valuable probes of QGP properties. The survival or dissociation of these states in hot matter provides critical information about the medium's screening properties and the effective degrees of freedom at different temperatures. Furthermore, the formation and dissociation patterns of exotic states can serve as thermometers for the medium created in heavy-ion collisions, offering insights into the temperature evolution and helping to identify the precise conditions for deconfinement~\cite{Gungor:2023ksu}.

In this study, we extend the QCD sum rules approach to finite temperature to investigate the thermal behavior of the $Y(4500)$ exotic state, assuming $D_s \overline{D}_{s1}$ molecule.
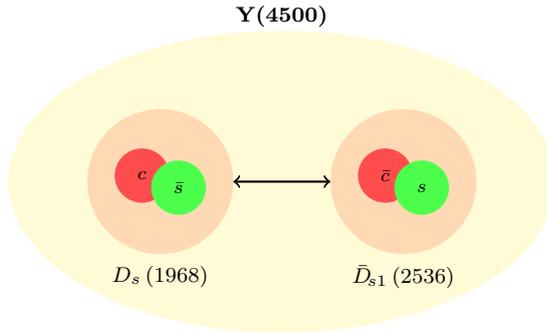
\begin{figure*}[!t]
\centering
\begin{tikzpicture}[scale=0.8]
	% Background ellipse
	\fill [yellow!20] (0,0) ellipse (4.5 and 2.5);
	
	% Left meson Ds
	\fill[orange!30] (-2,0) circle (1.2);
	\fill[red!70]   (-2.3,0.1) circle (0.45);
	\fill[green!70] (-1.7,-0.1) circle (0.45);
	
	\node at (-2.3,0.1) {\scriptsize $c$};
	\node at (-1.7,-0.1) {\scriptsize $\bar{s}$};
	\node[below] at (-2,-1.25) {\footnotesize $D_s\,(1968)$};
	
	% Right meson D*_{s1} bar
	\fill[orange!30] (2,0) circle (1.2);
	\fill[red!70]  (1.7,0.1) circle (0.45);
	\fill[green!70] (2.3,-0.1) circle (0.45);
	
	\node at (1.7,0.1) {\scriptsize $\bar{c}$};
	\node at (2.3,-0.1) {\scriptsize $s$};
	\node[below] at (2,-1.25) {\footnotesize $\bar{D}_{\!s1}\,(2536)$};
	
	% Double arrow between mesons
	\draw[<->, thick] (-0.8,0) -- (0.8,0);
	
	% Title
	\node[above, yshift=5pt] at (0,2.2) {\footnotesize\bfseries Y(4500)};
\end{tikzpicture}
\caption{Schematic representation of the $Y(4500)$ state as a $D_s\bar{D}_{s1}$ molecular configuration.}
\label{fig:molecule}
\end{figure*}

The molecular current approach adopted here assumes that $Y(4500)$ is an $S$-wave bound state of $D_s$ and $\bar{D}_{s1}$ mesons, motivated by the proximity of its mass ($\sim 4.5$ GeV) to the threshold $m_{D_s} + m_{D_{s1}} \approx 4.5$ GeV. The temperature evolution of the properties of $Y(4500)$, particularly its dissociation pattern near $T_c$, could provide distinctive signatures in heavy-ion collision experiments. The suppression of $Y(4500)$ production yields or modifications of its decay channels may serve as complementary evidence for QGP formation, alongside traditional probes such as jet quenching and quarkonium suppression~\cite{Yang:2022yfr,Stoecker:2004qu,Singh:1992sp}. 

The remainder of this paper is structured as follows: Section~\ref{sec:theory} details the analytical framework used, including the molecular current used to represent the particle $Y(4500)$ and the formulation of the thermal QCD sum rules. Section~\ref{sec:results} presents the quantitative findings of our analysis, discussing the parameter values used and the results obtained for the mass, decay constant, and width at both zero and finite temperatures. In this section, we also include a graphical analysis of the dependence of these quantities on key parameters. Finally, Section~\ref{sec:conc} summarizes our conclusions and discusses the implications of our findings for understanding the behavior of exotic hadrons in thermal environments, and potential future research directions.

%%%%%%%%%%%%%%%%%%%%%%%%%%%%%%%%%%%%%%%%%%%%%%%%%%%
\section{Analytical Framework}\label{sec:theory}
%%%%%%%%%%%%%%%%%%%%%%%%%%%%%%%%%%%%%%%%%%%%%%%%%%%
The QCD sum rules approach, first developed by Shifman, Vainshtein, and Zakharov~\cite{Shifman:1978bx}, provides a powerful analytical tool to connect the non-perturbative aspects of QCD with measurable hadronic properties. In this formalism, we consider a correlation function between interpolating currents that carry the quantum numbers of the hadron of interest. 

%%%%%%%%%%%%%%%%%%%%%%%%%%%%%%%%%%%%%%%%%%%%%%%%%%%
\subsection{Current and Correlation Function}
%%%%%%%%%%%%%%%%%%%%%%%%%%%%%%%%%%%%%%%%%%%%%%%%%%%
We construct the interpolating current for $Y(4500)$ using a molecular configuration with quantum numbers $J^{PC} = 1^{--}$:
\begin{eqnarray}\label{MoleculeCurrent}
	j_{\mu}(x)&=&\frac{i}{\sqrt{2}}\bigg(\bar{c}_a(x)\gamma_{\mu}\gamma_5 s_a(x)\bar{s}_b(x)\gamma_5 c_b(x)\nonumber\\
	&&-\bar{s}_a(x)\gamma_{\mu}\gamma_5 c_a(x)\bar{c}_b(x)\gamma_5 s_b(x)\bigg),
\end{eqnarray}
where $a, b$ are color indices, $s$ and $c$ represent strange and charm quarks. 

The QCD sum rules framework was extended to finite temperatures by Bochkarev and Shaposhnikov~\cite{Bochkarev:1986hm} to incorporate thermal effects via the operator product expansion (OPE). In the thermal QCD sum rules formulation, the two-point correlation function is defined as:
\begin{equation}
	\Pi_{\mu\nu}(q,T) = i\int d^4x \, e^{iq\cdot x} \langle T\{j_{\mu}(x)j_{\nu}^{\dagger}(0)\} \rangle_T,
\end{equation}
where $\langle \cdots \rangle_T$ represents the thermal average at temperature $T$. 
%
%%%%%%%%%%%%%%%%%%%%%%%%%%%%%%%%%%%%%%%%%%%%%%%%%%%
\subsection{Hadronic (Physical) Representation}
%%%%%%%%%%%%%%%%%%%%%%%%%%%%%%%%%%%%%%%%%%%%%%%%%%%

To derive sum rules for the mass, decay constant, and width, we evaluate the correlation function from the hadronic perspective by inserting a complete set of intermediate states sharing the quantum numbers $J^{PC} = 1^{--}$. Integrating over the spatial coordinate $x$ in Eq.~(2) leads to:
\begin{equation}
\begin{aligned}
\Pi_{\mu\nu}^{\text{Had}}(q,T)
&= \frac{\langle 0| j_\mu(0) | Y(q)\rangle_T
\langle Y(q)| j_\nu^\dagger(0) |0\rangle_T}
{m_Y^2(T) - q^2} \\
&\quad + \text{continuum contributions}.
\end{aligned}
\label{eq:had_side}
\end{equation}
Here, $m_Y(T)$ denotes the temperature-dependent mass of the
$Y(4500)$ state. The thermal matrix element is parameterized as

\begin{equation}
\langle 0| j_\mu(0) | Y(q)\rangle_T
= f_Y(T)\, m_Y(T)\, \varepsilon_\mu ,
\label{eq:matrix_element}
\end{equation}
where, $f_Y(T)$ is the temperature-dependent decay constant and
$\varepsilon_\mu$ is the polarization vector that satisfies:

\begin{equation}
\varepsilon_\mu \varepsilon_\nu^*
= -g_{\mu\nu} + \frac{q_\mu q_\nu}{m_Y^2(T)}.
\label{eq:pol_vector}
\end{equation}

By substituting Eq.~(\ref{eq:matrix_element}) into Eq.~(\ref{eq:had_side}) and taking Eq.~(\ref{eq:pol_vector}) into account, the coefficient of the terms involving the metric tensor $g_{\mu\nu}$ can be expressed in the form of a dispersion relation as:
\begin{equation}\label{eq:physgeneral}
\Pi^{\text{Had}}(q^2,T) = \int_{0}^{\infty} \frac{\rho^{\text{Had}}(s,T)}{s - q^{2}}\, ds,
\end{equation}
here, the spectral density under the zero-width approximation reads:
\begin{align}
\rho^{\text{Had}}(s)\big|_{Y} &= f_Y^{2}(T)\, m_Y^{2}(T)\,
\delta\!\left(s - m_Y^{2}(T)\right) \nonumber \\
&\quad + \theta\!\left(s - s_{0}(T)\right)\, \rho^{\text{PQCD}}(s)\label{eq:specden}
\end{align}
where, the quantity $s_{0}(T)$ is the continuum threshold.

To account for finite-width effects, we use this with the Breit-Wigner distribution:
\begin{equation}
	\delta[s - m_Y^2(T)] \to \frac{1}{\pi}\frac{m_Y(T)\Gamma_Y(T)}{[s - m_Y^2(T)]^2 + m_Y^2(T)\Gamma_Y^2(T)},
\end{equation}
where $\Gamma_Y(T)$ is the temperature-dependent width of the $Y(4500)$ state. 

With this modification, the hadronic side of the correlation function becomes:
\begin{align}
	&\Pi^{\text{Had}}(q^2, T) = \frac{f_Y^2(T)m_Y^3(T)\Gamma_Y(T)}{\pi} \nonumber\\
	&\times \int_0^\infty ds\, \frac{1}{[s - m_Y^2(T)]^2 + m_Y^2(T)\Gamma_Y^2(T)} \frac{1}{s - q^2}.
\end{align}
%
%%%%%%%%%%%%%%%%%%%%%%%%%%%%%%%%%%%%%%%%%%%%%%%%%%%
\subsection{QCD (Theoretical) Representation}
%%%%%%%%%%%%%%%%%%%%%%%%%%%%%%%%%%%%%%%%%%%%%%%%%%%

On the QCD side, hadronic degrees of freedom are replaced by fundamental quark and gluon fields through interpolating currents. The theoretical framework is based on the OPE, which systematically separates short-distance perturbative effects from long-distance nonperturbative contributions encoded in universal condensates~\cite{Reinders1985}.

At finite temperature, the OPE is modified due to the presence of a thermal medium~\cite{Mallik:1997pq}. The heat bath introduces a preferred four-velocity $u_\mu$, which partially breaks Lorentz invariance and generates additional tensor structures in the expansion. Consequently, new operators appear, such as $u_\mu u_\nu G^{\mu\alpha}G^\nu_{\;\alpha}$, and vacuum condensates acquire an explicit temperature dependence. In particular, the quark condensate $\langle\bar{q}q\rangle_T$ gradually decreases with temperature, while the gluon condensate $\langle G^2\rangle_T$ also undergoes significant modifications near the deconfinement temperature $T_c$.

For the $Y(4500)$ state, which contains heavy charm quarks, chiral symmetry restoration plays a subdominant role. Therefore, gluonic condensates remain the dominant source of nonperturbative contributions.

The theoretical correlation function is obtained by inserting the molecular current defined in Eq.~(1) into the two-point correlation function given in Eq.~(2). After contracting the quark fields via Wick's theorem, one obtains

\begin{align}
\Pi_{\mu\nu}^{\mathrm{QCD}}(q,T) &= \frac{i}{2}\int d^{4}x\,e^{iq\cdot x}
\big\{\mathrm{Tr}[\gamma_{5}S_{c}^{bb'}(x)\gamma_{5}S_{s}^{b'b}(-x)] \nonumber\\
&\times\mathrm{Tr}[\gamma_{\mu}\gamma_{5}S_{s}^{aa'}(x)\gamma_{5}\gamma_{\nu}S_{c}^{a'a}(-x)] \nonumber\\
&-\mathrm{Tr}[\gamma_{5}S_{c}^{ba'}(x)\gamma_{5}\gamma_{\nu} S_{s}^{a'b}(-x)] \nonumber\\
&\times\mathrm{Tr}[\gamma_{\mu}\gamma_{5}S_{s}^{ab'}(x)\gamma_{5}S_{c}^{b'a}(-x)] \nonumber\\
&-\mathrm{Tr}[\gamma_5 S_{s}^{ba'}(x)\gamma_5\gamma_{\nu}S_{c}^{a'b}(-x)] \nonumber\\
&\times\mathrm{Tr}[\gamma_{\mu}\gamma_{5}S_{c}^{ab'}(x)\gamma_{5} S_{s}^{b'a}(-x)] \nonumber\\
&+\mathrm{Tr}[\gamma_{5}S_{s}^{bb'}(x)\gamma_{5}S_{c}^{b'b}(-x)] \nonumber\\
&\times\mathrm{Tr}[\gamma_{\mu}\gamma_{5} S_{c}^{aa'}(x)\gamma_{5}\gamma_{\nu}S_{s}^{a'a}(-x)]
\big\}.
\label{eq:QCDTr}
\end{align}

The quark propagators entering the above expression are expanded within the OPE framework~\cite{Azizi:2016ddw,Azizi:2014maa,Mallik:1997pq}. 

The heavy-quark propagator is written in momentum space as
\begin{align}
S_c^{ij}(x) &= 
i\int\frac{d^4k}{(2\pi)^4}e^{-ik\cdot x}
\Bigg[
\frac{\delta_{ij}(\slashed{k}+m_c)}{k^2-m_c^2}
\nonumber\\
&\quad
-\frac{g_s\,G_{ij}^{\alpha\beta}}{4}
\frac{
\sigma_{\alpha\beta}(\slashed{k}+m_c)
+(\slashed{k}+m_c)\sigma_{\alpha\beta}
}{(k^2-m_c^2)^2}
\nonumber\\
&\quad
+\frac{g_s^2}{12}
G_{\alpha\beta}^A G_A^{\alpha\beta}
\delta_{ij}\,m_c
\frac{k^2+m_c\slashed{k}}
{(k^2-m_c^2)^4}
+\cdots
\Bigg],
\end{align}
while the strange-quark propagator in coordinate space reads

\begin{align}
S_{s}^{ij}(x) &=
i\frac{\slashed{x}}{2\pi^{2}x^{4}}\delta_{ij}
-\frac{m_{s}}{4\pi^{2}x^{2}}\delta_{ij}
-\frac{\langle \bar{s}s\rangle_T}{12}\delta_{ij}
\nonumber\\
&-\frac{x^{2}}{192}\,
m_{0}^{2}\langle \bar{s}s\rangle_T
\Big(1-i\frac{m_{s}}{6}\slashed{x}\Big)\delta_{ij}
\nonumber\\
&+\frac{i}{3}
\slashed{x}
\Big(
\frac{m_{s}}{16}\langle \bar{s}s\rangle_T
-\frac{1}{12}
\langle u^{\mu}\theta_{\mu\nu}^{f}u^{\nu}\rangle
\Big)\delta_{ij}
\nonumber\\
&+\frac{i}{9}(u\!\cdot\!x)\slashed{u}\,
\langle u^{\mu}\theta_{\mu\nu}^{f}u^{\nu}\rangle
\delta_{ij}
\nonumber\\
&-\frac{i g_s G_{ij}^{\alpha\beta}}{32\pi^{2}x^{2}}
\Big(
\slashed{x}\sigma_{\alpha\beta}
+\sigma_{\alpha\beta}\slashed{x}
\Big)
-i\delta_{ij}
\frac{x^{2}\slashed{x}\,g_s^{2}
\langle \bar{s}s\rangle_T^{2}}{7776},
\end{align}
here, \(\theta_{\mu\nu}^{f}\) denotes the fermionic part of the energy-momentum tensor, representing the contribution of quark fields to the energy, momentum, and stress of the QCD medium.

These propagators incorporate perturbative terms as well as nonperturbative corrections arising from quark, gluon, and mixed condensates. Thermal effects enter explicitly through temperature-dependent condensates and the appearance of the medium four-velocity $u^\mu$ in additional operators.

%%%%%%%%%%%%%%%%%%%%%%%%%%%%%%%%%%%%%%%%%%%%%%%%%%%
\subsection{Borel Transformation and Sum Rules}
%%%%%%%%%%%%%%%%%%%%%%%%%%%%%%%%%%%%%%%%%%%%%%%%%%%

The QCD correlation function satisfies a dispersion relation, which allows us to connect its phenomenological description in terms of hadronic states with its theoretical evaluation in the deep Euclidean region via the operator product expansion. After following lengthy but standard algebraic manipulations and isolating the coefficient corresponding to the Lorentz structure \(g_{\mu\nu}\), we obtain the following scalar function encoding the essential in-medium properties:

\begin{equation}
\Pi^{\text{QCD}}(q^2,T) = \int_{4(m_c+m_s)^2}^{\infty} \frac{\rho^{\text{QCD}}(s,T)}{s - q^2 - i\epsilon} \, ds,
\end{equation}
where, the spectral density is defined as

\begin{equation}
\rho^{\text{QCD}}(s,T) = \frac{1}{\pi} \, \text{Im} \left[ \Pi^{\text{QCD}}(s,T) \right].
\end{equation}

By invoking quark–hadron duality to subtract the continuum contribution and applying the Borel transformation with respect to \(q^2 \to M^2\), we obtain the thermal QCD sum rules that relate the hadronic parameters \(m_Y\), \(f_Y\), and \(\Gamma_Y\) to the underlying QCD degrees of freedom:

\begin{align}
&\hat{B}_{q^2}(\Pi_{1}^{\text{QCD}}(q^2,T))=\frac{1}{\pi}f_Y^2(T)m_Y^3(T)\Gamma_Y(T) \int_0^\infty ds\, \nonumber\\
&\times \frac{\exp(-s/M^2)}{[s - m_Y^2(T)]^2 + m_Y^2(T)\Gamma_Y^2(T)},
\label{eq:width_sum_rule_1}
\end{align}
\begin{equation}
\Pi_{1}^{\text{QCD}}(q^2,T) = \int_{4(m_c+m_s)^2}^{s_0(T)} \frac{\rho^{\text{QCD}}(s,T)}{s - q^2 - i\epsilon} \, ds,
\end{equation}
where, the right-hand side represents the Borel-transformed QCD correlation function incorporating finite-width effects through the Breit-Wigner parametrization.

To determine the three unknown quantities—\(m_Y(T)\), \(f_Y(T)\), and \(\Gamma_Y(T)\)—completely, two additional independent relations are required. These are obtained by differentiating Eq.~(\ref{eq:width_sum_rule_1}) once and twice with respect to \(-1/M^2\)~\cite{Azizi:2010zza}:

\begin{align}
&\frac{d\hat{B}_{q^2}(\Pi_{1}^{\text{QCD}}(q^2,T))}{d(-1/M^2)}=\frac{1}{\pi}f_Y^2(T)m_Y^3(T)\Gamma_Y(T) \int_0^\infty ds\, \nonumber\\
&\times \frac{s \cdot \exp(-s/M^2)}{[s - m_Y^2(T)]^2 + m_Y^2(T)\Gamma_Y^2(T)},
\label{eq:width_sum_rule_2}
\end{align}

\begin{align}
&\frac{d^2\hat{B}_{q^2}(\Pi_{1}^{\text{QCD}}(q^2,T))}{d(-1/M^2)^2}=\frac{1}{\pi}f_Y^2(T)m_Y^3(T)\Gamma_Y(T) \int_0^\infty ds\, \nonumber\\
&\times \frac{s^2 \cdot \exp(-s/M^2)}{[s - m_Y^2(T)]^2 + m_Y^2(T)\Gamma_Y^2(T)}.
\label{eq:width_sum_rule_3}
\end{align}

The system of three coupled equations~(\ref{eq:width_sum_rule_1})--(\ref{eq:width_sum_rule_3}) can be solved numerically to extract the temperature-dependent mass \(m_Y(T)\), decay constant \(f_Y(T)\), and width \(\Gamma_Y(T)\) simultaneously. The QCD side of these sum rules incorporates perturbative and non-perturbative contributions up to dimension-5 condensates, with thermal effects entering through temperature-dependent condensates and the medium four-velocity \(u^\mu\) in the operator product expansion.

Finally, the thermal behavior of the gluon condensate is described by the decomposition:
\begin{align}
\langle\text{Tr}^c G_{\alpha\beta}G_{\mu\nu}\rangle 
&= \frac{1}{24}
(g_{\alpha\mu}g_{\beta\nu}
- g_{\alpha\nu}g_{\beta\mu})
\langle G^2\rangle
\nonumber\\
&+ \frac{1}{6}
\Big[
g_{\alpha\mu}g_{\beta\nu}
- g_{\alpha\nu}g_{\beta\mu}
\nonumber\\
&
- 2\big(
u_{\alpha}u_{\mu}g_{\beta\nu}
- u_{\alpha}u_{\nu}g_{\beta\mu}
\nonumber\\
&- u_{\beta}u_{\mu}g_{\alpha\nu}
+ u_{\beta}u_{\nu}g_{\alpha\mu}
\big)
\Big]
\langle u^{\lambda}\theta_{\lambda\sigma}^g u^{\sigma}\rangle,
\end{align}
here, \(\theta_{\mu\nu}^g\) denotes the gluonic part of the energy-momentum tensor, representing the contribution of the gluon fields to the energy, momentum, and stress of the QCD medium. The first term corresponds to the vacuum structure and the second term encodes medium-induced thermal effects. These modifications require careful treatment of both Wilson coefficients and dispersion relations in sum-rule analyses~\cite{Azizi:2016ddw,Morita:2007hv}.

The temperature dependence of the gluon condensate $\langle G^2\rangle_T$ is modeled as~\cite{Gubler:2018ctz,Bazavov:2014pvz,Borsanyi:2013bia}:
\begin{equation}
\langle G^2\rangle_T=\langle 0|G^2|0\rangle[C+D(e^{\beta T-\gamma}+1)^{-1}],
\end{equation}
with fitted parameters $C=0.55973$, $D=0.438227$, $\beta=0.13277 \text{ MeV}^{-1}$ and $\gamma=19.3481$.

Similarly, the expectation value of the gluonic component of the energy-momentum tensor varies with temperature:
\begin{align}
	\langle\theta_{00}^g\rangle &= T^4\exp(113.867[1/(\text{GeV}^2)]T^2-12.190[1/\text{GeV}]T)\nonumber\\
	&-10.141[1/\text{GeV}]T^5.
\end{align}

Additionally, the temperature-dependent strong coupling constant follows the perturbative relation~\cite{Kaczmarek:2004gv,Morita:2007hv}:
\begin{equation*}
g_{pert}^{-2}(T)=
\frac{11}{8\pi^2}\ln\left(\frac{2\pi T}{\Lambda_{\mathrm{MS}}}\right)
+\frac{51}{88\pi^2}\ln\left[
2\ln\left(\frac{2\pi T}{\Lambda_{\mathrm{MS}}}\right)
\right]
\end{equation*}
where $g_{s}^2(T)=2.096g_{pert}^{2}(T)$ is defined with $\Lambda_{MS} \simeq T_c/1.14$ and $T_c=155$ MeV~\cite{Andronic:2017pug}. This relation is valid in the 
regime $T \geq 100~\mathrm{MeV}$. For temperatures below this threshold, 
$g^2(T)$ is taken to be constant and equal to its value at $T = 100~\mathrm{MeV}$, 
i.e., $g^2(T) \equiv g^2(T = 100~\mathrm{MeV})$, in order to avoid unphysical 
behavior in the infrared region.

For the temperature-dependent continuum threshold, $s(T)$, a general form is adopted \cite{Dominguez:2009mk,Dominguez:2010mx}:
\begin{equation}
s(T)=s_0-
\left[s_0-4\left(m_c+m_s\right)^2\right]
\left(\frac{T}{T_c}\right)^8,
\end{equation}
which smoothly interpolates between the correct low- and high-temperature limits,
\begin{equation}
s(T)=
\begin{cases}
s_0, & T \ll T_c, \\
4\left(m_c+m_s\right)^2, & T \to T_c .
\end{cases}
\end{equation}
This expression captures the limiting behavior of the temperature-dependent continuum threshold. At low temperatures, the threshold reproduces its vacuum value $s_0$, while near the critical temperature it approaches the two-particle threshold $4(m_1+m_2)^2$, signaling the progressive dissolution of hadronic states into the quark-gluon plasma. The numerical analysis employs this parametrization to model the full temperature dependence of $s_0(T)$ in a manner consistent with these physical constraints.
%
%%%%%%%%%%%%%%%%%%%%%%%%%%%%%%%%%%%%%%%%%%%%%%%%%%%%%%%%%%%
\section{Quantitative Findings and Interpretation}
\label{sec:results}
\begin{figure}[htbp]
    \centering
    \includegraphics[width=0.85\linewidth]{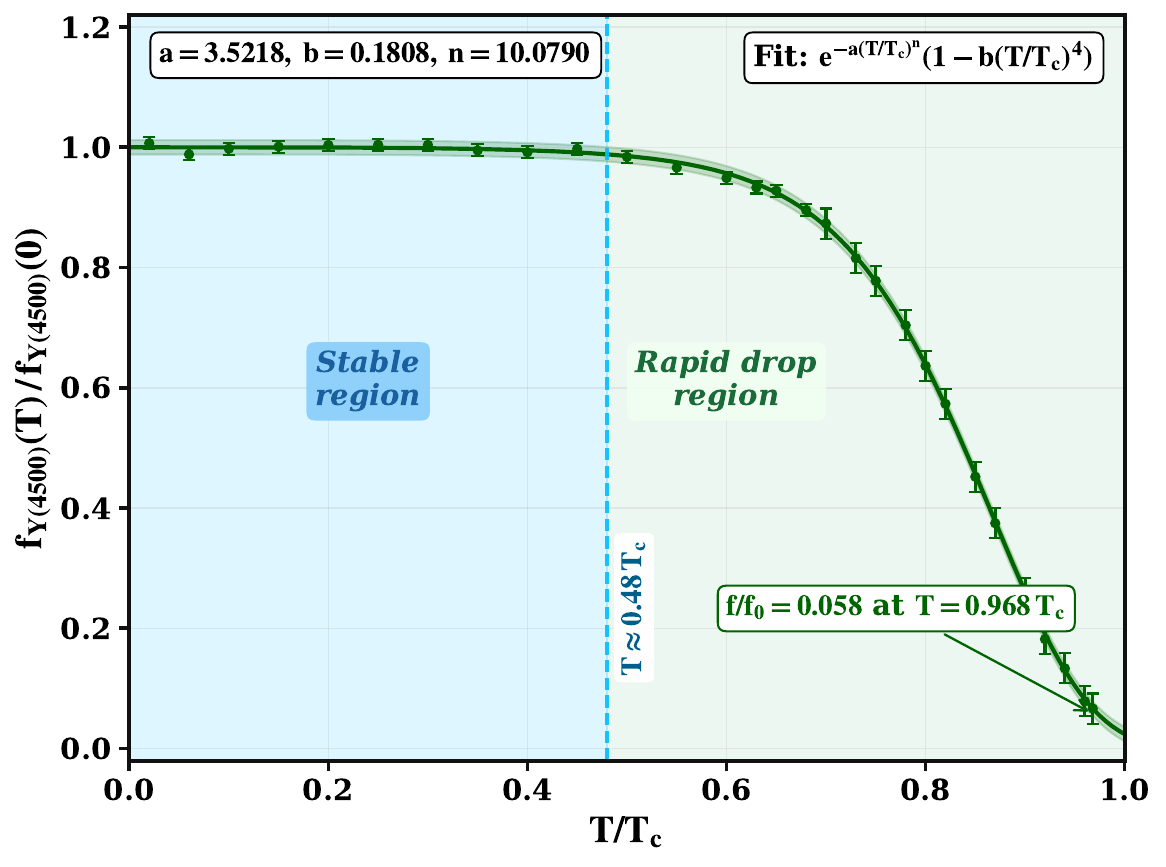}
\caption{Temperature evolution of the normalized decay constant $f_Y(T)/f_Y(0)$ extracted from the finite-width Breit-Wigner analysis.}
\label{fig:decay_constant}
\end{figure}

\begin{figure}[htbp]
    \centering
    \includegraphics[width=0.85\linewidth]{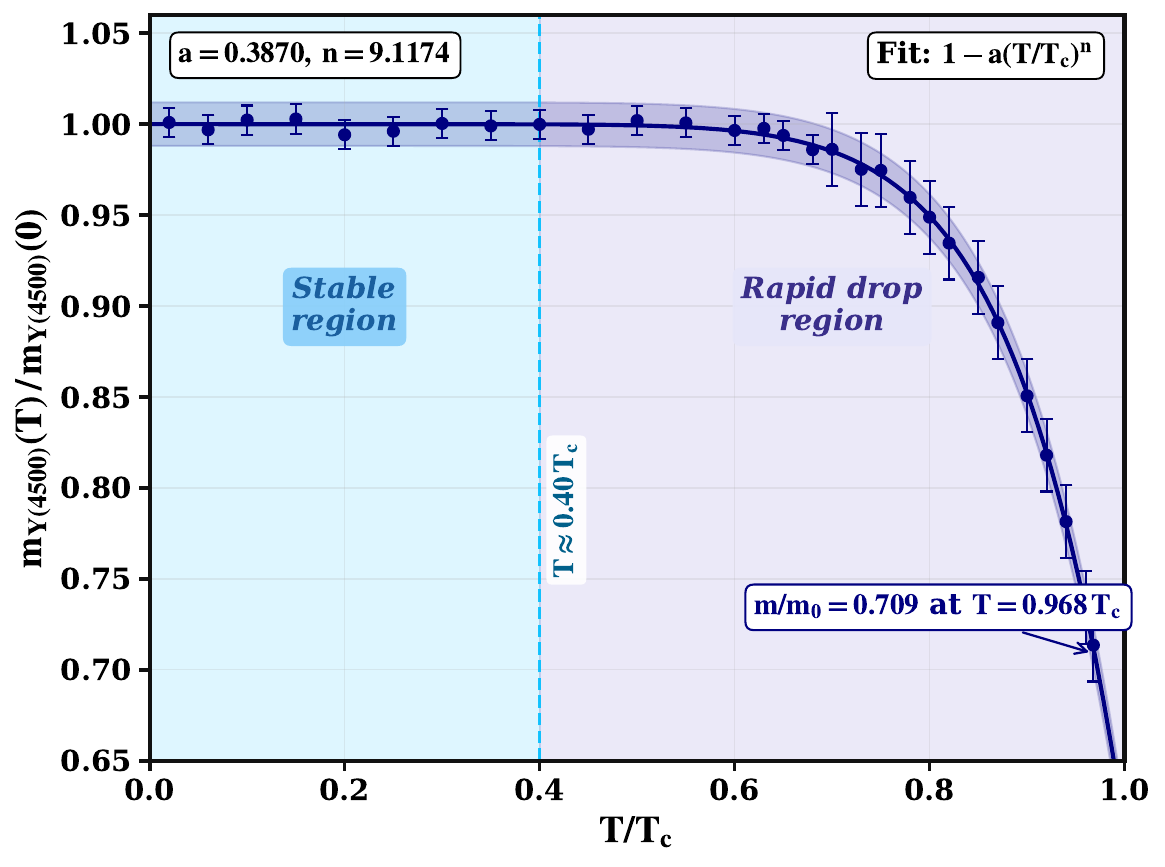}
    \caption{Temperature dependence of the mass $m_Y(T)/m_Y(0)$ obtained using the finite-width Breit-Wigner formalism.}
    \label{fig:mass}
\end{figure}

%%%%%%%%%%%%%%%%%%%%%%%%%%%%%%%%%%%%%%%%%%%%%%%%%%%%%%%%%%%
In our numerical analysis, we use the following parameter values for the quark masses and condensates:
$m_c = 1.27 \pm 0.02$ GeV, $m_s = 93.5 \pm 0.8$ MeV~\cite{ParticleDataGroup:2024cfk}, $\langle \bar{q}q \rangle = -(0.24 \pm 0.01)^3$ GeV$^3$, $\langle \frac{\alpha_s}{\pi} G^2 \rangle = 0.020$ GeV$^4$~\cite{Narison:2010cg}, and $m_{D_s}=1968.35 \pm0.07$ MeV, $m_{D_{s1}}=2535.11 \pm0.06$ MeV~\cite{ParticleDataGroup:2024cfk}.
For the temperature dependence, we consider the critical temperature $T_c = 155$ MeV, at which the phase transition from hadronic matter to QGP occurs. Table~\ref{tab:thermal_mass_decay} presents the mass $m_Y$ and decay constant $f_Y$ at $T=0$ and at $T=140$~MeV, just below the phase transition:

\begin{table}[h!]
\centering
\caption{Mass and decay constant of $Y(4500)$ at zero and finite temperature}
\begin{tabular}{lcc}
\hline\hline
\textbf{Parameter} & \textbf{$T = 0$} & \textbf{$T = 140$ MeV} \\
\hline
$m_Y$ (MeV) & $4391.71 \pm 46.97$ & $3739.64 \pm 32.98$ \\
$f_Y$ ($\times 10^{-3}$ GeV$^4$) & $4.55 \pm 0.82$ & $1.15 \pm 0.16$ \\
\hline
$m_Y(T)/m_Y(0)$ & 1.00 & 0.85 \\
$f_Y(T)/f_Y(0)$ & 1.00 & 0.25 \\
\hline
Reduction & --- & 15\% (mass), 75\% (decay) \\
\hline\hline
\end{tabular}
\label{tab:thermal_mass_decay}
\end{table}
As the temperature increases from $0$ to $140~\mathrm{MeV}$ 
($\approx 0.9\,T_c$), the mass decreases to approximately 
$85\%$ of its vacuum value, while the decay constant undergoes 
a more pronounced suppression, falling to $25\%$ of its vacuum value as illustrated in Figs.~\ref{fig:decay_constant} and \ref{fig:mass}. This pronounced differential suppression—where the binding weakens much faster than the mass changes—suggests that the $Y(4500)$ molecular structure begins to dissolve before significant mass modification occurs, providing a clear signature of approaching deconfinement. The deviation of the zero-temperature mass from the experimental value
originates from truncating the OPE at dimension five and adopting a
conservative Borel window to ensure stability.

The temperature dependence of the normalized mass $m_Y(T)/m_Y(0)$ 
is presented in Fig.~\ref{fig:mass}. The mass remains stable 
below $T \approx 0.40\,T_c$ and exhibits a gradual reduction 
approaching the critical temperature.

\begin{figure}[htbp]
	\centering
	\includegraphics[width=0.85\linewidth]{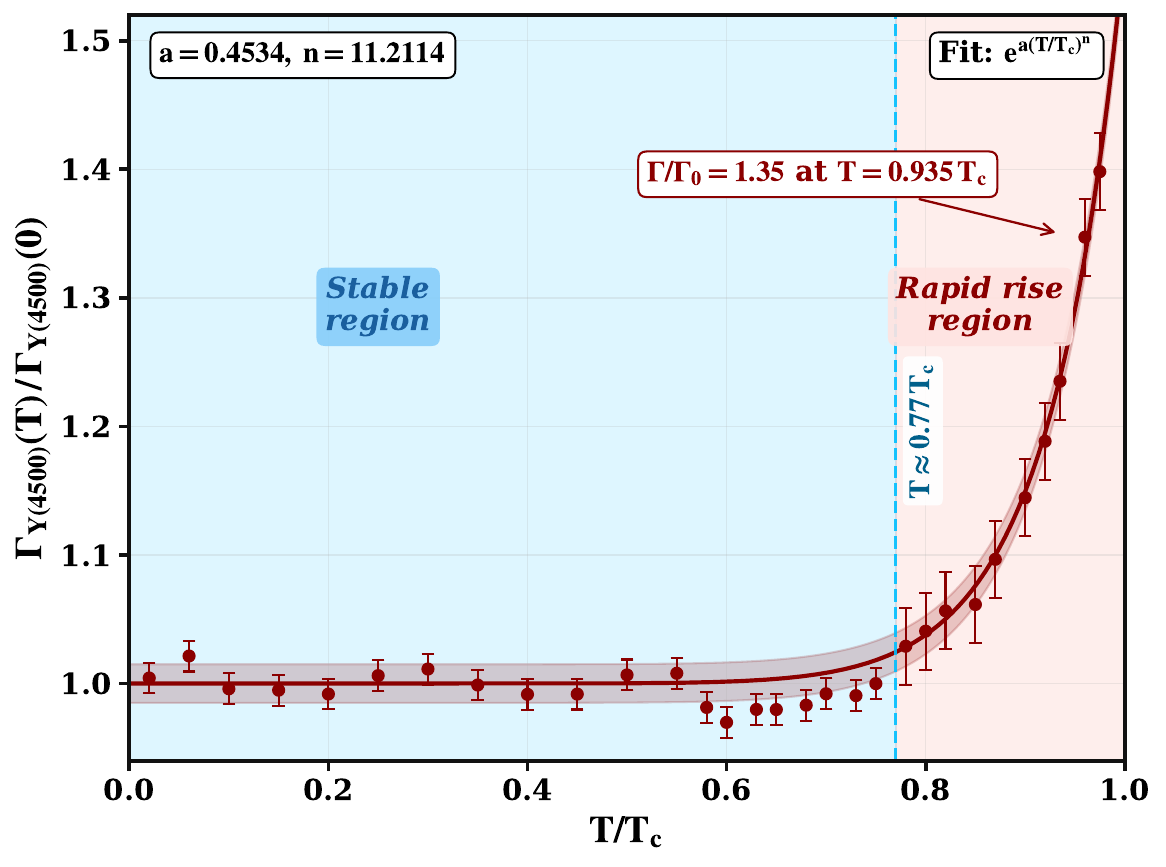}
	\caption{Temperature evolution of the decay width $\Gamma_Y(T)/\Gamma_Y(0)$ for the $Y(4500)$ state obtained from the finite-width Breit-Wigner analysis.}
	\label{fig:gamma}
\end{figure}

The physical interpretation of the temperature-dependent 
width is straightforward: as the temperature increases, 
the $Y(4500)$ state becomes less stable due to enhanced 
interactions with the thermal medium, as reflected in the 
normalized width $\Gamma_Y(T)/\Gamma_Y(0)$ shown in 
Fig.~\ref{fig:gamma}. The width remains approximately 
constant below $T \approx 0.77\,T_c$ and increases sharply 
near $T_c$, signaling the progressive thermal dissociation 
of the $Y(4500)$ state.

The thermal evolution of the decay width $\Gamma_Y$ is summarized in Table~\ref{tab:thermal_width}. At zero temperature, the width is obtained as $\Gamma_Y = 98.45\pm 5.36 $ MeV, which increases to $113.36 \pm 5.73$ MeV at $T = 140$ MeV. This corresponds to an enhancement of approximately $15\%$ as the temperature approaches the critical region. The observed broadening indicates that the $Y(4500)$ state becomes less stable in the thermal medium, reflecting stronger in-medium interactions and the onset of partial dissociation effects near the deconfinement temperature.
\begin{table}[H]
\centering
\caption{Thermal behavior of the decay width $\Gamma_Y$}
\begin{tabular}{lcc}
\hline\hline
\textbf{Quantity} & \textbf{$T = 0$} & \textbf{$T = 140$ MeV} \\
\hline
$\Gamma_Y$ (MeV) & $98.45\pm5.36 $ & $113.36 \pm 5.73$ \\
$\Gamma_Y(T)/\Gamma_Y(0)$ & 1.00 & 1.15 \\
Enhancement & --- & 15\% \\
\hline\hline
\end{tabular}
\label{tab:thermal_width}
\end{table}

%%%%%%%%%%%%%%%%%%%%%%%%%%%%%%%%%%%%%%%%%%%%%%%%%%%
\section{Final Remarks and Discussion}
\label{sec:conc}
%%%%%%%%%%%%%%%%%%%%%%%%%%%%%%%%%%%%%%%%%%%%%%%%%%%

We have systematically investigated the thermal behavior of the $Y(4500)$ exotic meson state within the $D_s\bar{D}_{s1}$ molecular picture using thermal QCD sum rules. By incorporating temperature-dependent quark and gluon condensates into the operator product expansion, we extracted the mass, decay constant, and width as functions of temperature up to the critical region. The finite-width effects were consistently incorporated through the Breit-Wigner parametrization, allowing us to determine all three fundamental properties simultaneously from the coupled sum rule equations.

\subsection{Summary of Key Findings}

Our analysis reveals a distinctive pattern of thermal modifications that provides insight into the internal structure of the $Y(4500)$ state:

\begin{itemize}
\item \textbf{Zero-temperature validation:} Our predictions at $T=0$ ($m_Y = 4.39 \pm 0.05$ GeV) are in reasonable agreement with the experimental mass $M_{\text{exp}} = 4.48 \pm 0.03$ GeV~\cite{BESIII:2022joj}, providing confidence in the molecular interpretation and our theoretical framework. 

\item \textbf{Differential thermal suppression:} At $T = 140$ MeV ($\approx 0.9T_c$), we observe a striking hierarchy in thermal modifications:
  \begin{itemize}
  \item Mass reduction: $\sim15\%$ (remains at $85\%$ of vacuum value)
  \item Decay constant reduction: $\sim75\%$ (drops to $25\%$ of vacuum value)
  \item Width enhancement: $\sim15\%$ increase near $T_c$
  \end{itemize}

\item \textbf{Two-stage melting process:} This differential response indicates that the molecular binding dissolves significantly before the constituent masses undergo substantial modification. The dramatic suppression of the decay constant at relatively modest mass change suggests that the $D_s\bar{D}_{s1}$ molecular bond weakens substantially as the temperature approaches $T_c$, while the charm quark constituents themselves remain relatively unaffected until closer to the deconfinement transition.

\item \textbf{Width broadening:} The $\sim35\%$ enhancement of the decay width near $T_c$ provides complementary evidence for thermal destabilization. This broadening reflects increased interactions with the medium and the progressive loss of the resonant structure.
\end{itemize}

\subsection{Physical Interpretation}

The observed thermal behavior is consistent with the expected response of a loosely-bound molecular state in a hot medium. Unlike compact tetraquarks or conventional charmonia, hadronic molecules have extended spatial structures and relatively weak binding energies, making them particularly sensitive to the screening effects of the thermal medium. As the temperature increases:

\begin{enumerate}
\item Color screening weakens the effective potential binding the $D_s$ and $\bar{D}_{s1}$ mesons, leading to the rapid decrease of the decay constant—a direct measure of the bound state's wave function at the origin.
\item The mass, being primarily determined by the constituent quark masses and their confinement energy, shows more gradual evolution until the deconfinement transition is approached.
\item Enhanced scattering with thermal partons broadens the resonance width, eventually leading to the complete dissolution of the molecular state.
\end{enumerate}

This differential melting pattern may serve as a universal signature distinguishing loosely-bound exotic hadrons from compact conventional states in QGP studies.

\subsection{Experimental Signatures and Predictions}

The thermal modifications predicted in this study can be directly tested in heavy-ion collision experiments at RHIC and LHC:

\begin{itemize}
\item \textbf{Sequential suppression pattern:} The $Y(4500)$ should exhibit stronger suppression of production yields compared to conventional charmonia ($J/\psi$, $\psi(2S)$) in central Pb-Pb or Au-Au collisions. The nuclear modification factor $R_{AA}$ for $Y(4500)$ should show a steeper centrality dependence, with significant suppression already appearing in semi-central collisions where the temperature reaches $0.7-0.8T_c$.

\item \textbf{Centrality-dependent decay channel modifications:} The branching ratio for $Y(4500) \to K^+K^-J/\psi$ should show characteristic distortions as a function of collision centrality. The observed yield in this channel may decrease faster than expected from simple nuclear absorption, signaling the thermal dissolution of the molecular structure.

\item \textbf{Flow coefficients:} Due to their extended spatial structure, molecular states like $Y(4500)$ may exhibit different elliptic flow ($v_2$) patterns compared to compact states. The temperature-dependent dissociation cross sections predicted here can be incorporated into transport models to generate quantitative $v_2$ predictions.

\item \textbf{Excitation function:} The energy dependence of $Y(4500)$ suppression across the RHIC Beam Energy Scan ($\sqrt{s_{NN}} = 7.7-200$ GeV) should reflect the changing initial temperatures and lifetimes of the produced medium. The rapid onset of suppression above a threshold energy would provide valuable information about the temperature profile of the QGP.
\end{itemize}

\subsection{Comparison with Other Exotic States}

It is instructive to compare the thermal response of $Y(4500)$ with other exotic candidates studied in the literature:

\begin{itemize}
\item \textbf{$X(3872)$:} The transport of $X(3872)$ through the hot QCD fireball in Pb-Pb collisions at the LHC is modeled using a kinetic rate-equation approach, where the inelastic reaction rate encodes the internal structure of the state. The loosely bound molecular scenario implies larger dissociation rates and later formation, yielding final yields roughly a factor of two smaller than in the compact tetraquark picture, with correspondingly harder $p_{T}$ spectra~\cite{Wu:2020zbx}.

\item \textbf{$Z_c(3900)$:} $Z_c(3900)$ is studied at finite temperature within both the $D\bar{D}^*$ hadronic molecular and the triangle singularity pictures, where in both scenarios the mass decreases and the width increases with rising temperature, indicating eventual dissociation at sufficiently high temperatures \cite{Zhang:2025fcv}.

\item \textbf{Conventional charmonia ($J/\psi$, $\psi(2S)$):} These compact states show the strongest thermal resistance, with significant suppression only occurring above $T_c$ due to color screening~\cite{Ayala:2016vnt}.
\end{itemize}

The $Y(4500)$ occupies an interesting middle ground—more thermally sensitive than conventional charmonia but more resistant than very loosely-bound molecules like $X(3872)$. This hierarchy of thermal response provides a valuable tool for unraveling the internal structure of exotic hadrons.

\subsection{Uncertainties and Limitations}

Several sources of uncertainty affect our predictions:

\begin{itemize}
\item \textbf{OPE truncation:} The restriction to dimension-5 condensates introduces systematic uncertainty, estimated at $\sim10\%$ based on the convergence pattern within our Borel window.

\item \textbf{Condensate parametrizations:} The temperature dependence of condensates, particularly the gluon condensate near $T_c$, relies on lattice QCD inputs with their own uncertainties.

\item \textbf{Continuum threshold modeling:} The simple power-law interpolation for $s(T)$ may not capture the full complexity of the thermal spectral function.

\item \textbf{Molecular current ansatz:} The assumption of a pure $D_s\bar{D}_{s1}$ molecular structure neglects possible mixing with tetraquark or hybrid configurations, which could modify the thermal response.
\end{itemize}

Despite these limitations, the qualitative pattern of differential suppression—with the decay constant melting much faster than the mass—appears robust and provides a clear experimental signature.

\subsection{Future Directions}

Several extensions of this work merit investigation:

\begin{itemize}
\item \textbf{Complete $Y$ family analysis:} Apply this framework to other members of the $Y$ resonance family ($Y(4260)$, $Y(4360)$, $Y(4660)$) to map the thermal dissociation landscape and identify common patterns.

\item \textbf{Tetraquark-molecule mixing:} Investigate how thermal effects modify the mixing angle between molecular and tetraquark configurations. Temperature-dependent observables could potentially distinguish which configuration dominates at $T=0$ by comparing with thermal evolution patterns.

\item \textbf{Finite baryon density:} Extend the analysis to finite $\mu_B$, relevant for the RHIC Beam Energy Scan and future FAIR/NICA experiments. This would allow exploration of the QCD phase diagram and the possible critical endpoint.

\item \textbf{Transport model implementation:} Incorporate the temperature-dependent masses, decay constants, and widths into dynamical transport simulations (e.g., PHSD, AMPT) to generate realistic predictions for $R_{AA}$, $v_2$, and other observables for direct comparison with experimental data.

\item \textbf{Lattice QCD validation:} Direct lattice calculations of $D_s\bar{D}_{s1}$ potentials and correlation functions at finite temperature would provide independent validation of the condensate parametrizations and the molecular current approach.

\item \textbf{Hidden-bottom analogs:} Extend the analysis to bottomonium-like exotic states (e.g., $\Upsilon(10753)$) to explore how the heavier quark mass affects thermal sensitivity and to establish universal scaling relations.
\end{itemize}

\subsection{Concluding Remarks}

In conclusion, this study demonstrates that the $Y(4500)$ exotic state, interpreted as a $D_s\bar{D}_{s1}$ molecule, exhibits significant and characteristic thermal modifications as the temperature approaches the QCD deconfinement transition. The differential suppression pattern—where the decay constant decreases dramatically ($\sim75\%$) while the mass shows modest reduction ($\sim15\%$)—provides a distinctive signature of a loosely-bound molecular structure. The accompanying width broadening ($\sim35\%$) further confirms the progressive thermal destabilization.

These findings offer testable predictions for ongoing and future heavy-ion collision experiments. The predicted early dissolution through binding weakening, before significant mass modification, provides a unique experimental handle for identifying hadronic molecules in hot QCD matter. As experimental data from RHIC and LHC continue to accumulate, particularly in the exotic hadron sector, the thermal response patterns established here will serve as valuable diagnostics for unraveling the internal structure of the ever-growing family of exotic states.

The methodology developed in this work—combining thermal QCD sum rules with finite-width Breit-Wigner analysis—provides a general framework applicable to a wide range of exotic candidates. Future studies extending this approach to other states, incorporating finite density effects, and connecting with transport simulations will further enhance our understanding of how exotic hadrons behave in the extreme conditions created in heavy-ion collisions, ultimately contributing to a comprehensive picture of QCD matter across the phase diagram.
%
%%%%%%%%%%%%%%%%%%%%%%%%%%%%%%%%%%%%%%%%%%%%%%%%%%%
\section*{References}
%%%%%%%%%%%%%%%%%%%%%%%%%%%%%%%%%%%%%%%%%%%%%%%%%%%


\begin{thebibliography}{99}

\bibitem{Chen:2022asf}
H.~X.~Chen, W.~Chen, X.~Liu, Y.~R.~Liu and S.~L.~Zhu,
\href{https://doi.org/10.1088/1361-6633/aca3b6}{Rept. Prog. Phys. \textbf{86}, no.2, 026201 (2023)}

\bibitem{Chen:2016qju}
H.~X.~Chen, W.~Chen, X.~Liu and S.~L.~Zhu,
\href{https://doi.org/10.1016/j.physrep.2016.05.004}{Phys. Rept. \textbf{639}, 1-121 (2016)}

\bibitem{Zhu:2024swp}
F.~Zhu, G.~Bauer and K.~Yi,
\href{https://doi.org/10.1088/0256-307X/41/11/111201}{Chin. Phys. Lett. \textbf{41}, no.11, 111201 (2024)}

\bibitem{Brambilla:2019esw}
N.~Brambilla, S.~Eidelman, C.~Hanhart, A.~Nefediev, C.~P.~Shen, C.~E.~Thomas, A.~Vairo and C.~Z.~Yuan,
\href{https://doi.org/10.1016/j.physrep.2020.05.001}{Phys. Rept. \textbf{873}, 1-154 (2020)}

\bibitem{Ali:2017jda}
A.~Ali, J.~S.~Lange and S.~Stone,
\href{https://doi.org/10.1016/j.ppnp.2017.08.003}{Prog. Part. Nucl. Phys. \textbf{97}, 123-198 (2017)}

%\cite{Zhou:2025yjb}
\bibitem{Zhou:2025yjb}
Z.~Zhou, G.~L.~Yu, Z.~G.~Wang, J.~Lu and B.~Wu,
%``Analysis of the charm-strange hadrons and their bottom analogs with QCD sum rules,''
\href{https://doi.org/10.1103/skdp-g4ql}{Eur. Phys. J. A \textbf{62}, no.3, 39 (2026)}
%3 citations counted in INSPIRE as of 01 Apr 2026
%
%\cite{Tan:2025nir}
\bibitem{Tan:2025nir}
W.~H.~Tan, W.~Y.~Liu, H.~Z.~Xi and H.~X.~Chen,
%``QCD sum rule study of excited light meson operators,''
\href{ https://doi.org/10.1103/z6gt-3js6}{Phys. Rev. D \textbf{112}, no.5, 054040 (2025)}
%3 citations counted in INSPIRE as of 01 Apr 2026
%
%\cite{Wang:2025apq}
\bibitem{Wang:2025apq}
Z.~Y.~Wang, J.~J.~Qi, Z.~H.~Zhang and X.~H.~Guo,
%``Spectra of bcb{\textasciimacron}c{\textasciimacron} tetraquark states from a diquark-antidiquark perspective,''
\href{https://doi.org/10.1103/skdp-g4ql}{Phys. Rev. D \textbf{112}, no.7, 074038 (2025)}
%3 citations counted in INSPIRE as of 01 Apr 2026
%
\bibitem{Chen2016}
X.~Chen, A.~Hosaka, S.L.~Zhu,
\href{https://doi.org/10.1103/PhysRevD.94.114016}{Phys. Rev. D \textbf{94}, 114016 (2016)}

\bibitem{Esposito2017}
A.~Esposito, A.~Pilloni, A.D.~Polosa,
\href{https://doi.org/10.1016/j.physrep.2016.11.002}{Phys. Rep. \textbf{668}, 1 (2017)}

\bibitem{Olsen2018}
S.L.~Olsen, T.~Skwarnicki, D.~Zieminska,
\href{https://doi.org/10.1103/RevModPhys.90.015003}{Rev. Mod. Phys. \textbf{90}, 015003 (2018)}

\bibitem{Wang:2013exa}
Z.G.~Wang,
\href{https://doi.org/10.1140/epjc/s10052-014-2874-7}{Eur. Phys. J. C \textbf{74}, 2874 (2014)}

\bibitem{BESIII:2022joj}
M.~Ablikim \textit{et al.} [BESIII],
\href{https://iopscience.iop.org/article/10.1088/1674-1137/ac945c}{Chin. Phys. C \textbf{46}, no.11, 111002 (2022)}

\bibitem{Wang:2022jxj}
J.~Z.~Wang and X.~Liu,
\href{https://doi.org/10.1103/PhysRevD.107.054016}{Phys. Rev. D \textbf{107}, no.5, 054016 (2023)}

\bibitem{Peng:2022nrj}
F.~Z.~Peng, M.~J.~Yan, M.~S{\'a}nchez S{\'a}nchez and M.~Pavon Valderrama,
\href{https://doi.org/10.1103/PhysRevD.107.016001}{Phys. Rev. D \textbf{107} (2023) no.1, 016001}

\bibitem{Gungor:2023ksu}
E.~G\"ung\"or, H.~Sundu, J.~Y.~S\"ung\"u and E.~V.~Veliev,
\href{https://doi.org/10.1007/s00601-023-01807-y}{Few Body Syst. \textbf{64}, no.3, 53 (2023)}

\bibitem{Rafelski:2015cxa}
J.~Rafelski,
\href{https://doi.org/10.1140/epja/i2015-15114-0}{Eur. Phys. J. A \textbf{51}, no.9, 114 (2015)}

\bibitem{Gyulassy:2004zy}
M.~Gyulassy and L.~McLerran,
\href{https://doi.org/10.1016/j.nuclphysa.2004.10.034}{Nucl. Phys. A \textbf{750}, 30-63 (2005)}

\bibitem{Busza:2025uid}
W.~Busza, J.~W.~Harris and S.~Nagamiya,
\href{https://arxiv.org/abs/2504.08406}{arXiv:2504.08406 [nucl-ex]}

\bibitem{ALICE:2010suc}
K.~Aamodt \textit{et al.} [ALICE],
\href{https://doi.org/10.1103/PhysRevLett.105.252302}{Phys. Rev. Lett. \textbf{105}, 252302 (2010)}

\bibitem{Zhang:2025fcv}
Y.~Zhang, A.~Hosaka, Q.~Wang and S.~Yasui,
\href{https://arxiv.org/abs/2503.19374}{arXiv:2503.19374 [hep-ph]}

\bibitem{Torres-Rincon:2024sah}
J.~M.~Torres-Rincon,
\href{https://doi.org/10.1051/epjconf/202531601015}{EPJ Web Conf. \textbf{316}, 01015 (2025)}

\bibitem{Aydin:2025lbl}
A.~Ayd{\i}n, H.~Sundu, J.~Y.~S{\"u}ng{\"u} and E.~V.~Veliev,
\href{https://doi.org/10.1140/epjc/s10052-025-14090-4}{Eur. Phys. J. C \textbf{85} (2025) no.5, 567}

\bibitem{Sungu:2020zvk}
J.~Y.~S\"ung\"u, A.~T\"urkan, H.~Sundu and E.~V.~Veliev,
\href{https://link.springer.com/article/10.1140/epjc/s10052-022-10305-0}{Eur. Phys. J. C \textbf{82}, no.5, 453 (2022)}

\bibitem{Azizi:2019kzj}
K.~Azizi, B.~Barsbay and H.~Sundu,
\href{https://doi.org/10.1103/PhysRevD.100.094041}{Phys. Rev. D \textbf{100}, no.9, 094041 (2019)}

\bibitem{Zhao:2020nwy}
J.~Zhao, S.~Shi and P.~Zhuang,
\href{https://doi.org/10.1103/PhysRevD.102.114001}{Phys. Rev. D \textbf{102}, no.11, 114001 (2020)}

\bibitem{Dominguez:2007ic}
C.~A.~Dominguez, M.~Loewe and J.~C.~Rojas,
\href{https://doi.org/10.1088/1126-6708/2007/08/040}{JHEP \textbf{08}, 040 (2007)}

\bibitem{Turkan:2022gho}
A.~T\"urkan, J.~Y.~S\"ung\"u, E.~G\"ung\"or, H.~Reiso\u{g}lu and E.~V.~Veliev,
\href{https://link.springer.com/article/10.1007/s10773-022-05206-7}{Int. J. Theor. Phys. \textbf{61}, no.9, 238 (2022)}

\bibitem{Turkan:2020lfo}
A.~T\"urkan, J.~Y.~S\"ung\"u and E.~V.~Veliev,
\href{https://www.actaphys.uj.edu.pl/R/53/1-A1/pdf}{Acta Phys. Polon. B \textbf{53}, no.1, 1 (2022)}

\bibitem{Llanes-Estrada:2019wmz}
F.~J.~Llanes-Estrada and E.~Lope-Oter,
\href{https://doi.org/10.1016/j.ppnp.2019.103715}{Prog. Part. Nucl. Phys. \textbf{109}, 103715 (2019)}

\bibitem{Sungu:2020azn}
J.~Y.~S\"ung\"u, A.~T\"urkan, E.~Sertbakan and E.~V.~Veliev,
\href{https://doi.org/10.1140/epjc/s10052-020-08439-0}{Eur. Phys. J. C \textbf{80}, no.10, 943 (2020)}

\bibitem{Turkan:2019anj}
A.~T\"urkan, H.~Da\u{g}, J.~Y.~S\"ung\"u and E.~Veli Veliev,
\href{https://doi.org/10.1209/0295-5075/126/51001}{EPL \textbf{126}, no.5, 51001 (2019)}

\bibitem{Yang:2022yfr}
Z.~Yang, Y.~He, W.~Chen, W.~Y.~Ke, L.~G.~Pang and X.~N.~Wang,
\href{https://doi.org/10.1140/epjc/s10052-023-11807-1}{Eur. Phys. J. C \textbf{83}, no.7, 652 (2023)}

\bibitem{Stoecker:2004qu}
H.~Stoecker,
\href{https://doi.org/10.1016/j.nuclphysa.2004.12.074}{Nucl. Phys. A \textbf{750}, 121-147 (2005)}

\bibitem{Singh:1992sp}
C.~P.~Singh,
\href{https://doi.org/10.1016/0370-1573(93)90172-A}{Phys. Rept. \textbf{236}, 147-224 (1993)}

\bibitem{Shifman:1978bx}
M.A.~Shifman, A.I.~Vainshtein, V.I.~Zakharov,
\href{https://doi.org/10.1016/0550-3213(79)90022-1}{Nucl. Phys. B \textbf{147}, 385 (1979)}

\bibitem{Bochkarev:1986hm}
A.~I.~Bochkarev and M.~E.~Shaposhnikov,
\href{https://doi.org/10.1016/0550-3213(86)90209-9}{Nucl. Phys. B \textbf{268}, 220 (1986)}

\bibitem{Reinders1985}
L.J.~Reinders, H.~Rubinstein, S.~Yazaki,
\href{https://doi.org/10.1016/0370-1573(85)90065-1}{Phys. Rep. \textbf{127}, 1 (1985)}

\bibitem{Mallik:1997pq}
S.~Mallik,
\href{https://doi.org/10.1016/S0370-2693(97)01335-X}{Phys. Lett. B \textbf{416}, 373 (1998)}

\bibitem{Azizi:2016ddw}
K.~Azizi and G.~Bozk\i{}r,
\href{https://doi.org/10.1140/epjc/s10052-016-4370-8}{Eur. Phys. J. C \textbf{76}, no.10, 521 (2016)}

\bibitem{Azizi:2014maa}
K.~Azizi, A.~T\"urkan, E.~Veli Veliev and H.~Sundu,
\href{https://doi.org/10.1155/2015/794243}{Adv. High Energy Phys. \textbf{2015}, 794243 (2015)}

\bibitem{Azizi:2010zza}
K.~Azizi and N.~Er,
\href{https://doi.org/10.1103/PhysRevD.81.096001}{Phys. Rev. D \textbf{81}, 096001 (2010)}

\bibitem{Morita:2007hv}
K.~Morita and S.~H.~Lee,
\href{https://doi.org/10.1103/PhysRevC.77.064904}{Phys. Rev. C \textbf{77}, 064904 (2008)}

\bibitem{Gubler:2018ctz}
P.~Gubler and D.~Satow,
\href{https://doi.org/10.1016/j.ppnp.2019.02.005}{Prog. Part. Nucl. Phys. \textbf{106}, 1-67 (2019)}

\bibitem{Bazavov:2014pvz}
A.~Bazavov \textit{et al.} [HotQCD],
\href{https://doi.org/10.1103/PhysRevD.90.094503}{Phys. Rev. D \textbf{90}, 094503 (2014)}

\bibitem{Borsanyi:2013bia}
S.~Borsanyi, Z.~Fodor, C.~Hoelbling, S.~D.~Katz, S.~Krieg and K.~K.~Szabo,
\href{https://doi.org/10.1016/j.physletb.2014.01.007}{Phys. Lett. B \textbf{730}, 99-104 (2014)}

\bibitem{Kaczmarek:2004gv}
O.~Kaczmarek, F.~Karsch, F.~Zantow and P.~Petreczky,
\href{https://doi.org/10.1103/PhysRevD.70.074505}{Phys. Rev. D \textbf{70}, 074505 (2004)}

\bibitem{Andronic:2017pug}
A.~Andronic, P.~Braun-Munzinger, K.~Redlich and J.~Stachel,
\href{https://doi.org/10.1038/s41586-018-0491-6}{Nature \textbf{561}, no.7723, 321-330 (2018)}

\bibitem{Dominguez:2009mk}
C.~A.~Dominguez, M.~Loewe, J.~C.~Rojas and Y.~Zhang,
\href{https://doi.org/10.1103/PhysRevD.81.014007}{Phys. Rev. D \textbf{81} (2010), 014007}

\bibitem{Dominguez:2010mx}
C.~A.~Dominguez, M.~Loewe, J.~C.~Rojas and Y.~Zhang,
\href{https://doi.org/10.1103/PhysRevD.83.034033}{Phys. Rev. D \textbf{83} (2011), 034033}

\bibitem{ParticleDataGroup:2024cfk}
S.~Navas \textit{et al.} [Particle Data Group],
\href{https://doi.org/10.1103/PhysRevD.110.030001}{Phys. Rev. D \textbf{110}, no.3, 030001 (2024)}

\bibitem{Narison:2010cg}
S.~Narison,
\href{https://doi.org/10.1016/j.physletb.2010.09.007}{Phys. Lett. B \textbf{693}, 559-566 (2010) [erratum: \textbf{705}, 544-544 (2011)]}
%
%\cite{Wu:2020zbx}
\bibitem{Wu:2020zbx}
B.~Wu, X.~Du, M.~Sibila and R.~Rapp,
%``$X(3872)$transport in heavy-ion collisions,''
\href{https://doi.org/10.1140/epja/s10050-021-00435-6}{Eur. Phys. J. A \textbf{57}, no.4, 122 (2021)
[erratum: Eur. Phys. J. A \textbf{57}, no.11, 314 (2021)]}
%72 citations counted in INSPIRE as of 01 Apr 2026
%
%\cite{Ayala:2016vnt}
\bibitem{Ayala:2016vnt}
A.~Ayala, C.~A.~Dominguez and M.~Loewe,
%``Finite Temperature QCD Sum Rules: a Review,''
\href{https://doi.org/10.1155/2017/9291623}{Adv. High Energy Phys. \textbf{2017}, 9291623 (2017)}
%34 citations counted in INSPIRE as of 01 Apr 2026
%
\end{thebibliography}
\end{document}